\documentclass[12pt]{article}
\usepackage{mathrsfs}

\def\gsim{\mathop{\smash{>}}\limits_\sim}

\title{\bf Gauged $\mbox{\boldmath $B-3L_\tau$}$, low-energy
unification and proton decay} 

\author{
\bf Palash B. Pal\footnote{pbpal@theory.saha.ernet.in}\\
Saha Institute of Nuclear Physics\\
1/AF Bidhan-Nagar, Calcutta 700064, India \\
\and
\bf Utpal Sarkar\footnote{utpal@prl.ernet.in}\\
Physical Research Laboratory, Ahmedabad 380 009
}

\date{}

\begin{document}
\maketitle

\begin{abstract}\noindent
We point out that if there is a gauged $B-3L_\tau$ symmetry
at low energy, it can prevent fast proton decay. This may
help building models with theories with extra dimensions at the
TeV scale. For purpose of illustration we present an explicit
model with large extra dimensions. The Higgs required for a 
realistic fermion masses and mixing are included. The problem
of neutrino masses are solved with triplet Higgs scalars. The
proton remains stable even after the $B-3L_\tau$ symmetry
breaking.

\end{abstract}

The minimal standard model of electroweak and strong interactions is
based on the gauge group $\rm SU(3)_c \times SU(2)_L\times
U(1)_Y$. Apart from this gauge symmetry, it has accidental global
symmetries, which are:
\begin{list}{\alph{enumi})}{\usecounter{enumi}\itemsep=0pt}
\item the baryon number $B$;
\item the generational lepton numbers $L_A$, where $A$ stands for the
charged leptons $e$, $\mu$ and $\tau$.
\end{list}
These symmetries are anomalous.  A common strategy for venturing to
gauge theories beyond the standard model is to gauge some linear
combination of these symmetries which is anomaly-free. Thus we have
obtained models with gauged $B-L$ quantum number \cite{Pati:1974yy},
which is the only non-anomalous global symmetry of the standard model
that treats all fermion generations on the same footing. In addition,
there are also ``non-universal'' models where the generational
universality is violated by gauging quantities of the form
$L_A-L_B$.

In the recent literature, a different type of non-anomalous global
symmetries of the standard model and the prospects of gauging such
symmetries has been studied \cite{Ma:1997nq}.  These are symmetries of
the form $B-3L_A$. For phenomenological reasons, the $B-3L_\tau$
symmetry has been considered most seriously and a simple model with a
U(1) symmetry based on this quantum number has been constructed
\cite{Ma:1997nq, Ma:1998dr, Ma:1998dp}. We point out that in models of
low energy unification this symmetry can play an important role.

In recent times an interesting possibility is emerging, in which
it is possible to have a TeV scale unification. 
It has been proposed \cite{Arkani-Hamed:1998rs} that if
there are compact extra dimensions with very large radius, in which
only gravity propagates and the ordinary particles are confined in the
usual $3+1$ space-time dimensions, then it is natural to have a very
low effective Planck scale. In these theories gravity will get unified
with other gauge interactions at a few TeV scale. All the gauge
interactions, namely, the strong and the electroweak interactions will
also get unified at a few TeV.  There is another class of theories
\cite{Randall:1999ee} with small extra dimensions, where the extra
dimensions are warped, giving rise to a low energy unification. The
hierarchy problem disappears in this latter class of theories.

In any realistic model with low unification scale, there are a few
generic problems which need to be solved.  The most important of them
are the reasons for the astounding stability of the proton and the
minuteness of neutrino masses.  In 4-dimensional unification models,
both proton decay rate and neutrino masses are suppressed by inverse
powers of a high scale, e.g., the unification scale.  This requires
large mass scales in the theory, $10^{15}$\,GeV or higher.  Low energy
unification models do not have such high scales.  The purpose of the
present article is to show how these two problems may be addressed in
the context of a low energy unification model.  We point out that in
the non-universal gauge model with a $B-3L_\tau$ symmetry, it is
possible to prevent proton decay altogether, without any constraint on
the possible unification scale, and without resorting to any mechanism
originating from the extra dimensions.  Neutrino mass, on the other
hand, is suppressed because of intricacies of a higher dimensional
theory, as we will show later.

We first discuss proton stability.  The simplest baryon number
breaking operators allowed by the gauge symmetry and the fermion
content of the standard model have dimension six
\cite{Weinberg:sa,Weinberg:1980bf}.  In 4-dimensional spacetime, they
are therefore suppressed by the inverse square of some large mass
scale $M$.  Consequently, one needs $M \gsim 10^{15}$\,GeV so that
proton lifetime becomes large enough to be consistent with
experiments.  In theories with large extra dimensions, $M$ is of the
order of a few TeV, and so one must find a mechanism to prevent fast
proton decay.

There are a few suggestions to ensure proton stability at levels
beyond predicted by the above argument.  For example, in 4-dimensional
unification models, it was shown that if baryon number is a part of
the gauge symmetry in the unified model, the effective operators for
proton decay have very high dimensions and so the scale can be much
lower \cite{Pal:1993yv}. Other suggestions involve physics from extra
dimensions \cite{exprot}.

We now discuss what happens in the model at hand.  To begin with, we
write the representations of all known fermion fields under $\rm
SU(2)_L\times U(1)_Y\times U(1)_{B-3L_\tau}$.  We omit the $\rm
SU(3)_c$ representation for the sake of brevity.  The index $i$ runs
over all generations of fermions, whereas $a$ runs only over the first
two.
\begin{eqnarray}
&q_{iL} = (2,\frac16,\frac13), \quad  u_{iR} = (1,\frac23,\frac13),
\quad  d_{iR} = (1,-\frac13,\frac13), \nonumber\\*
&\psi_{aL} = (2,-\frac12,0), \quad  \ell_{aR} = (1,-1,0), \nonumber\\*
&\psi_{\tau L} =  (2,-\frac12,-3), \quad  \tau_R = (1,-1,-3) .
\label{fermions}
\end{eqnarray}
%

\begin{table}[t]
\caption{Scalar multiplets with integral $B-3L_\tau$ which
can couple to fermions.  We have not included couplings involving
right-handed neutrinos.\label{t:higgs}}
\bigskip
\begin{center}
\begin{tabular}{|c|c|l|}
\hline
Name & Representation & Yukawa couplings \\
\hline
$\phi$ & (2,$\frac12$,0) \vphantom{$\frac11^{\frac11}$}
& $\overline q_{iL} d_{jR}$, $\overline
\psi_{aL} \ell_{bR}$, $\overline \psi_{\tau L} \tau_R$ \\
$\widetilde\phi$ & (2,$-\frac12$,0) & $\overline q_{iL} u_{jR}$\\
\hline
$\xi_0$ & (3,1,0) & $\psi_{aL} \psi_{bL}$ \\
$\xi_3$ & (3,1,3) & $\psi_{aL} \psi_{\tau L}$ \\
$\xi_6$ & (3,1,6) & $\psi_{\tau L} \psi_{\tau L}$ \\
\hline
$\chi^+_0$ & (1,1,0) & $\psi_{aL} \psi_{bL}, \qquad (a \neq b)$ \\
$\chi^+_3$ & (1,1,3) & $\psi_{aL} \psi_{\tau L}$ \\
\hline
$L^{++}_0$ & (1,2,0) & $\ell_{aR} \ell_{bR}$ \\
$L^{++}_3$ & (1,2,3) & $\ell_{aR} \tau_R$ \\
$L^{++}_6$ & (1,2,6) & $\tau_R \tau_R$ \\
\hline
\end{tabular}
\end{center}
\end{table}

In this model anomaly cancellation requires introduction of
right-handed neutrinos. In models with $B-L$ gauged symmetry
\cite{Pati:1974yy} all three generations of right-handed neutrinos are
required.  In the gauged $B-3L_\tau$ models, only the right-handed
tau-neutrino, $\nu_{\tau R}$, is necessary in this regard.  However,
to keep the analysis general, one may also add $\nu_{eR}$ and
$\nu_{\mu R}$, which will be gauge singlets.  Thus the list shown in
Eq.\ (\ref{fermions}) should be augmented by the following ones in
order to obtain all fermions in the model:
\begin{eqnarray}
\nu_{eR} = (1,0,0),  \qquad 
\nu_{\mu R} = (1,0,0), \qquad 
\nu_{\tau R} = (1,0,-3)\,.
\end{eqnarray}

With this fermion content, it will be possible to discuss the question
of baryon number violation by looking into the effective higher
dimensional operators.  We perform this analysis in two steps.  In the
first step, we disregard the effects of ${\rm U(1)}_{B-3L_\tau}$
breaking and discuss the nature of baryon number violation in the
theory with unbroken ${\rm U(1)}_{B-3L_\tau}$ symmetry.  In the next
stage, effects of breaking of this symmetry is brought in.

Any operator which breaks baryon number should have at least three
quarks, so the minimal operator must contain the combination $\cal
QQQ$, where $\cal Q$ is any quark multiplet, left handed or right
handed.  This combination has $B-3L_\tau=1$ and is not consistent
with the symmetry. If we add leptons, we can change the $(B-3L_\tau)$
values in units of three and can never form an invariant. The lowest
dimensional $(B-3L_\tau)$ invariant operator with non-zero baryon
number is
\begin{eqnarray}
{\cal Q}^9 L_\tau \,,
\label{Q9L}
\end{eqnarray}
where $L_\tau$ is a leptonic multiplet of the third generation, i.e.,
any one of the set $\psi_{\tau L}$, $\tau_R$ and $\nu_{\tau R}$.  This
operator has a coefficient $1/M^{11}$, where $M$ is the scale of new
physics.  This provides a huge suppression for any baryon number
violating process.  Further, this breaks baryon number by 3 units, and
therefore cannot induce proton decay.  $|\Delta B|=1$ operators are
impossible to construct, as commented above, and so proton is
completely stable in this model.  Simplest baryon number violating
processes arising out of the operator in Eq.\ (\ref{Q9L}) are, for
example,
\begin{eqnarray}
n + n &\to& \bar n + \bar \nu_\tau \,, \nonumber\\*
n + p &\to& \bar n + \bar \tau^+ \,.
\end{eqnarray}

To understand if the proton remains stable even after the gauged
$B-3L_\tau$ is broken, let us include the Higgs scalars in our
discussion. As we will discuss later, the neutrino masses and mixing
in theories of extra dimensions may require a few different Higgs
multiplets beyond the standard model Higgs multiplet $\phi$. So we
shall demonstrate the stability of proton with a rather generalized
Higgs content.

The standard model Higgs multiplet will be neutral under $\rm
U(1)_{B-3L_\tau}$. If our model contained only this Higgs multiplet,
it would not have been any different from the standard model, except
that the $\tau$ leptons would not have interacted with other leptons,
which might have led to some inconsistency. In order to break ${\rm
U(1)}_{B-3L_\tau}$, we need some other Higgs multiplets.  We restrict
ourselves to multiplets which have integral $B-3L_\tau$ quantum
numbers, and which can couple to fermions. Table~\ref{t:higgs} gives a
list of all such multiplets, along with fermion bilinears with which
they can couple.  Clearly, if we restrict ourselves to scalar
multiplets shown in this table, there will be no proton decay even
after all symmetry breaking.  The reason is simple.  Baryon number
will remain an accidental symmetry of the model even after introducing
all Yukawa couplings, and it will not be broken by any vacuum
expectation value of scalars.

Let us now discuss a scenario of fermion masses and mixing in this
gauged $B-3L_\tau$ model, including the question of neutrino masses.
The quark masses and mixing and the charged lepton masses come from
the Yukawa interactions with the usual standard model Higgs scalar
$\phi$,
\begin{eqnarray}
\mathscr L_Y &=& \sum_{i,j} f^{(u)}_{ij} \bar q_{iL} u_{jR} \phi 
+ \sum_{i,j} f^{(d)}_{ij} \bar q_{iL} d_{jR}
\phi^\dagger + \Big[ \sum_{a,b} f^{(\ell)}_{ab} \overline
\psi_{aL} \ell_{bR} + f_{\tau\tau} 
\overline \psi_{\tau L} \tau_R \Big] \phi^\dagger \nonumber \\
&& + \Big[ \sum_{a,b} f^{(\nu)}_{ab} \overline
\psi_{aL} \nu_{\ell bR} + f^{\nu}_{\tau\tau} 
\overline \psi_{\tau L} \nu_{\tau R} \Big] \phi
+ \mbox {h.c.} \,.
\end{eqnarray}
The up and down quark masses and mixing are not constrained in the
model with this simplest choice of Higgs scalar.  In the leptonic
sector, the charged current mixing matrix will depend on the neutrino
mass matrix. For a realistic neutrino mass matrix with appropriate
mixing, the $B-3L_\tau$ symmetry has to be broken and the Higgs
scalars should transform non-trivially under $B-3L_\tau$.

To solve the problem of hierarchy between the three generations of
fermions, we consider the thick wall scenario \cite{thick}.  The basic
assumption is that the particles in our brane are not confined to a
point.  Instead, they have a Gaussian profile.  The Gaussian width for
the scalar particle is much larger than all other fermions, so that
all the fermions have complete overlap with the scalar. However, the
overlaps between the profiles of left and right-handed particles are
not maximal. This would then introduce a suppression factor, which can
then explain the hierarchy between the three generations of fermions.

Consider a five dimensional example, with co-ordinates $z = \{ x,y
\}$, where $x$ stands for the 4-dimensional co-ordinates.  For
simplicity we assume that the scalar field $\phi$ is same over the
entire thick wall and falls off sharply outside the wall, but any
fermion $\Psi$ has a Gaussian profile in the extra dimension $y$,
centered around $y_0$:
\begin{eqnarray}
\Psi(z) = A e^{- \mu^2 (y - y_0)^2} \psi (x) ,
\end{eqnarray}
where $\psi$ is a normalized four-dimensional massless left-handed
fermion; $A= (2 \mu^2/\pi)^{1/4}$ is the normalization and $\mu$ is
the slope of the profile.  A generic Yukawa coupling term in the
action $\mathscr A$ will be
\begin{eqnarray}
\mathscr A_Y = \int d^5 z \; \sqrt{L} \Lambda \overline\Psi_i \Psi_j
\Phi = \int d^4 x \lambda \overline\psi_i \psi_j \phi,
\end{eqnarray}
where $L$ is the wall thickness and the four-dimensional Yukawa
coupling is given by
\begin{eqnarray}
\lambda = \int dy \; \Lambda A e^{- \mu^2 (y - y_i)^2} 
A e^{- \mu^2 (y - y_j)^2} = \Lambda e^{- \mu^2 (y_i - y_j)^2/2} .
\end{eqnarray}
The only assumption here is that wall thickness $L$ is much
larger than the Gaussian width $\mu^{-1}$. 

Depending on the positions $y_i$ and $y_j$ of the left and
right handed particles, the four dimensional Yukawa couplings 
could then have a hierarchy. A large hierarchy between the three
generations of fermions could
thus be generated by localizing the fields at a small distance in
units of $\mu^{-1}$, within the thick wall. 

Since the present results on neutrinos indicate only three
Majorana neutrinos and preferably no sterile neutrinos, we 
decouple the right-handed neutrinos from our discussion using 
the same thick wall mechanism. We assume that the profile of the left and
the right-handed neutrinos have very little overlap and hence the 
Dirac mass terms are almost zero. Later we introduce a 
Majorana mass term for the right-handed neutrinos at the scale
of $B-3 L_\tau$ breaking, which makes the right-handed Majorana
neutrinos to be heavy with very little mixing with the 
left-handed neutrinos. For the rest of the discussions
we shall thus ignore the right-handed neutrinos and their 
couplings with the left-handed neutrinos. For completeness,
we shall now show how left-handed neutrinos could have small
masses in this model.

The smallness of the neutrino masses is usually explained
by introducing a large lepton number violating scale $M_L$ in the
denominator and allowing an effective higher dimensional
operator which allows a Majorana mass of the neutrino. 
In theories with TeV scale gravity, the neutrino masses and
mixing requires new physics since there are no large lepton
number violating scale in the theory. There are a few suggestions
to solve this problem and we shall be discussing here one
such solutions in some detail~\cite{trip}. 

We work in a model of large extra dimensions, in which our four
dimensional world (3-brane ${\cal P}$ at $y=0$) is localized at
the origin of a higher ($n> 4$) dimensional space. The usual
fermions, gauge bosons and the Higgs doublets propagate only
in our 3-brane and are blind to all extra dimensions, while only
gravity propagates in the higher dimensional bulk. The extra
dimensions are compact and have a large radius $r$, but since
the overlap of the gravitons with the particles in our brane is
extremely small, the scale of gravity could be as low as a few
TeV. For phenomenological reasons we assume the scale of gravity
and compactification of the extra dimension is about 100 TeV,
which is the fundamental scale in the theory.

We include triplet Higgs scalars with different $B-3L_\tau$
charges, $\xi_0, \xi_3$ and $\xi_6$ in our 3-brane, which couples
to the leptons
\begin{eqnarray}
\mathscr L_\xi = \sum_{a,b} f_{ab} \xi_0 \psi_a \psi_b 
+ \sum_a f_{a \tau} \xi_3 \psi_a \psi_\tau 
+ f_{\tau \tau} \xi_6 \psi_\tau \psi_\tau +
\mbox{h.c.} \,, 
\end{eqnarray}
as seen in Table~\ref{t:higgs}. The scalar potential for these
fields $\xi_i, i=0,3,6$ are so chosen that they do not acquire
any vacuum expectation value (VEV) even when the $B-3L_\tau$
symmetry is broken. This will ensure that total lepton number 
$L = L_e + L_\mu + L_\tau$ 
is conserved in our brane. If there are no interactions of these
fields with any bulk matter, then the total lepton number $L$ will not
be broken at any time.

We consider one extra $SU(2)_L$ doublet
scalar $\kappa \equiv (2,{1 \over 2},-3)$. The Yukawa coupling of this
field is given by
\begin{eqnarray}
\mathscr L = \sum_{a} f^{(\nu)}_{a \tau} \overline
\psi_{aL} \nu_{\tau R} \kappa
+ \mbox {h.c.} \,.
\end{eqnarray}
This will determine the total lepton number of $\kappa$ to be
$L=0$. Since $L$ is conserved in our brane, the interactions
of the type $\xi \phi \phi$, $\xi_3 \kappa \phi$ and
$\xi_6 \kappa \kappa$ are all absent. All trilinear interactions
of the triplet Higgs scalars will now be forbidden by the 
total lepton number $L$ conservation in our brane. 

We now introduce a singlet scalar
\begin{eqnarray}
\sigma \equiv (1,0,0) 
\end{eqnarray}
which can propagate in the bulk.  This singlet carries lepton number
$L=2$, but it does not have any gauge interactions.  In the present
scenario the total lepton number $L$ is broken in another distant
3-brane (${\cal P}'$ at $ y=y_*$), when the singlet scalar
\begin{eqnarray}
\eta \equiv (1,0,0) 
\end{eqnarray}
acquires a VEV. This scalar $\eta$ carries lepton number $L=2$. Due to
the interaction of this field $\eta$ with the bulk scalar $\sigma$,
the lepton number violation in the brane ${\cal P}'$ will be
communicated to our brane through the interaction of the bulk scalar
$\sigma$ with fields in our brane.

Lepton number is conserved everywhere before the field $\eta$ acquires
$vev$.  The lepton number conserving interactions of the field
$\sigma$ are given by,
\begin{eqnarray}
S_\sigma &=& \int_{\cal P} d^4 x \bigg[ h_1
\xi_0^\dagger (x) \phi(x) \phi(x) \sigma (x, y=0) +
h_2
\xi_3^\dagger (x) \phi(x) \kappa(x) \sigma (x, y=0) 
\nonumber \\* 
&& +
h_3 \xi_6^\dagger (x) \kappa(x) \kappa(x) \sigma (x, y=0)
\bigg] +
\int_{{\cal P}'} d^4 x' \mu^2 ~\eta^\dagger(x')
\sigma (x', y=y_*) .
\end{eqnarray}
Lepton number violation from ${\cal P}'$ is transmitted to our brane
${\cal P}$ through the shined values of $\langle\sigma\rangle$,
\begin{eqnarray}
\langle \sigma (x, y=0) \rangle &=& \Delta_n (r) \langle \eta
(x,y=y_*) \rangle \,,
\end{eqnarray}
where $\langle\eta\rangle$ acts as point source and $\Delta_n(r)$ is
the Yukawa potential.  For an interesting choice with $m_\sigma r \ll
1$ and $n = 2$ the asymptotic form of the profile of $\sigma$
is~\cite{trip}
\begin{eqnarray}
\langle \sigma \rangle \approx {\Gamma({n-2 \over 2}) \over 4
\pi^{n/2}} {M_* \over (M_* r)^{n-2}} .
\end{eqnarray}
This will then induce an effective $vev$ for the triplet fields
$\xi(x)$ in our brane,
\begin{eqnarray}
\langle \xi_0 \rangle = h_1 {\langle \sigma \rangle \langle
\phi \rangle \langle \phi \rangle \over M_{\xi_0}^2} \,,\nonumber \\
\langle \xi_3 \rangle = h_2 {\langle \sigma \rangle \langle
\phi \rangle \langle \kappa \rangle \over M_{\xi_3}^2} \,,\nonumber \\
\langle \xi_6 \rangle = h_3 {\langle \sigma \rangle \langle
\kappa \rangle \langle \kappa \rangle \over M_{\xi_6}^2} \,.
\end{eqnarray}
The resultant neutrino mass matrix comes out to be
\begin{eqnarray}
{\cal M}_\nu = \pmatrix{ f_{ab}\langle \xi_0 \rangle & f_{a
\tau} \langle \xi_3 \rangle \cr f_{a \tau} \langle \xi_3 \rangle
& f_{\tau \tau} \langle \xi_6 \rangle }
\end{eqnarray}
with $a,b = e, \mu$. The Majoron in this scenario is a singlet and
its couplings to the charged fermions is suppressed by a factor
$m_\nu \sqrt{G_F}$. For appropriate choice of parameters this
model could explain the required neutrino masses and mixing.

The above scenario of large extra dimensions with the $B- 3
L_\tau$ gauge symmetry can now explain the neutrino masses without
any large scales in the theory, in addition to explaining the
quark and charged lepton masses and mixing. We have now introduced
three triplet Higgs, which carry $B-3L_\tau$ quantum numbers $0,3$
and $6$. This again ensures that there is no proton decay in the
theory. No new physics is required from extra dimensions to
prevent fast proton decay if the $B-3L_\tau$ symmetry is gauged, as
argued earlier in the paper.

We now turn to another possibility with supersymmetry and R-parity
violation. R-parity is usually imposed to prevent fast proton
decay in models with supersymmetry. In models with gauged
$B-3L_\tau$ symmetry all baryon number violating and R-parity
violating renormalizable terms are all absent. As a result, in
supersymmetric models with gauged $B-3L_\tau$ symmetry, there is
no need to impose R-parity. Proton decay is naturally suppressed.

In conclusion, we point out that in models with gauged $B-3L_\tau$
symmetry, proton decay is automatically suppressed. We constructed
one model of extra large dimensions with gauged $B-3L_\tau$ and
without any large scale in the theory, which can explain the
neutrino mass problem. Proton decay is suppressed without any
additional input. In R-parity violating supersymmetric models also
there is no problem of proton decay, if the $B-3L_\tau$ symmetry
is gauged.

We thank Gautam Bhattacharyya for discussions.



\end{document}